\documentclass[prl,twocolumn,superscriptaddress]{revtex4-1}
\usepackage{graphicx}
\usepackage{amsmath,graphicx}
\usepackage{epsfig}
\usepackage{amsmath,amssymb,mathrsfs}
\usepackage[bookmarks=true,
   colorlinks=true,
   linkcolor=blue,
   urlcolor=blue,
   citecolor=blue,
   bookmarks=true,
   hyperindex=true
]{hyperref}
\usepackage{natbib}

\usepackage{color}

\def\be{\begin{equation}}
\def\ee{\end{equation}}
\def\bea{\begin{eqnarray}}
\def\eea{\end{eqnarray}}

\newcommand{\tabincell}[2]
{\begin{tabular}{@{}#1@{}}#2\end{tabular}}

\begin{document}

\title{Interaction induced decay of a heteronuclear two-atom system}

\author{Peng Xu}
\affiliation{State Key Laboratory of Magnetic Resonance and Atomic and Molecular Physics, Wuhan Institute of Physics and Mathematics, Chinese Academy of Sciences - Wuhan National Laboratory for Optoelectronics, Wuhan 430071, China}
\affiliation{Center for Cold Atom Physics, Chinese Academy of Sciences, Wuhan 430071, China}

\author{Jiaheng Yang}
\affiliation{State Key Laboratory of Magnetic Resonance and Atomic and Molecular Physics, Wuhan Institute of Physics and Mathematics, Chinese Academy of Sciences - Wuhan National Laboratory for Optoelectronics, Wuhan 430071, China}
\affiliation{Center for Cold Atom Physics, Chinese Academy of Sciences, Wuhan 430071, China}
\affiliation{University of Chinese Academy of Sciences, Beijing 100049, China}

\author{Min Liu}
\affiliation{State Key Laboratory of Magnetic Resonance and Atomic and Molecular Physics, Wuhan Institute of Physics and Mathematics, Chinese Academy of Sciences - Wuhan National Laboratory for Optoelectronics, Wuhan 430071, China}
\affiliation{Center for Cold Atom Physics, Chinese Academy of Sciences, Wuhan 430071, China}

\author{Xiaodong He}
\affiliation{State Key Laboratory of Magnetic Resonance and Atomic and Molecular Physics, Wuhan Institute of Physics and Mathematics, Chinese Academy of Sciences - Wuhan National Laboratory for Optoelectronics, Wuhan 430071, China}
\affiliation{Center for Cold Atom Physics, Chinese Academy of Sciences, Wuhan 430071, China}

\author{Yong Zeng}
\affiliation{State Key Laboratory of Magnetic Resonance and Atomic and Molecular Physics, Wuhan Institute of Physics and Mathematics, Chinese Academy of Sciences - Wuhan National Laboratory for Optoelectronics, Wuhan 430071, China}
\affiliation{Center for Cold Atom Physics, Chinese Academy of Sciences, Wuhan 430071, China}
\affiliation{University of Chinese Academy of Sciences, Beijing 100049, China}

\author{Kunpeng Wang}
\affiliation{State Key Laboratory of Magnetic Resonance and Atomic and Molecular Physics, Wuhan Institute of Physics and Mathematics, Chinese Academy of Sciences - Wuhan National Laboratory for Optoelectronics, Wuhan 430071, China}
\affiliation{Center for Cold Atom Physics, Chinese Academy of Sciences, Wuhan 430071, China}
\affiliation{University of Chinese Academy of Sciences, Beijing 100049, China}

\author{Jin Wang}
\affiliation{State Key Laboratory of Magnetic Resonance and Atomic and Molecular Physics, Wuhan Institute of Physics and Mathematics, Chinese Academy of Sciences - Wuhan National Laboratory for Optoelectronics, Wuhan 430071, China}
\affiliation{Center for Cold Atom Physics, Chinese Academy of Sciences, Wuhan 430071, China}

\author{D.J. Papoular}
\affiliation{INO-CNR BEC Center and Dipartimento di Fisica, Universit\`a di Trento, 38123 Povo, Italy}

\author{G.V. Shlyapnikov}
\affiliation{State Key Laboratory of Magnetic Resonance and Atomic and Molecular Physics, Wuhan Institute of Physics and Mathematics, Chinese Academy of Sciences - Wuhan National Laboratory for Optoelectronics, Wuhan 430071, China}
\affiliation{Laboratoire de Physique Th\'eorique et Mod\`eles Statistiques, Universit\'e Paris Sud, CNRS, Orsay, France}
\affiliation{\mbox{Van der Waals-Zeeman Institute, University of Amsterdam, Science Park 904, 1098 XH Amsterdam, The Netherlands}}
\affiliation{Russian Quantum Center, Novaya Street 100, Skolkovo, Moscow Region 143025, Russia}

\author{Mingsheng Zhan}
\email{mszhan@wipm.ac.cn}
\affiliation{State Key Laboratory of Magnetic Resonance and Atomic and Molecular Physics, Wuhan Institute of Physics and Mathematics, Chinese Academy of Sciences - Wuhan National Laboratory for Optoelectronics, Wuhan 430071, China}
\affiliation{Center for Cold Atom Physics, Chinese Academy of Sciences, Wuhan 430071, China}

\date{\today}


\begin{abstract}
Two-atom systems in small traps are of fundamental interest, first of all for understanding the role of interactions in degenerate cold gases and for the creation of quantum gates in quantum information
processing with single-atom traps. One of the key quantities is the inelastic relaxation (decay) time when one of the atoms or both are in a higher hyperfine state. Here we measure this quantity in a heteronuclear
system of $^{87}$Rb and $^{85}$Rb in a micro optical trap and demonstrate experimentally and theoretically the presence of both fast and slow relaxation processes, depending on the choice of the initial hyperfine
states. The developed experimental method allows us to single out a particular relaxation process and, in this sense, our experiment is a "superclean platform" for collisional physics studies. Our results have
also implications for engineering of quantum states via controlled collisions and creation of two-qubit quantum gates.
\end{abstract}


\maketitle

\subsection*{Introduction}
The studies of two-atom systems in small traps attract a great deal of interest, in particular for engineering of quantum states via controlled collisions and creation of quantum gates in quantum information processing with a set of single-atom traps \cite{Jaksch1999,Mandel2003}. The crucial points are the decoherence time and the lifetime related to the interaction-induced inelastic decay of a higher hyperfine state. On the other hand, this type of inelastic processes, in particular heteronuclear ones, are important for the creation of multi-species quantum degenerate systems \cite{multi-degenerate}, for obtaining ultracold heteronuclear molecules \cite{mole-review}, and for ultracold chemistry \cite{chemistry}. However, the main obstacle in the studies of inelastic heteronuclear collisions in a trapped gas \cite{Collision-review} is a simultaneous presence of a large variety of loss mechanisms, which complicates the analysis. In magneto-optical traps where many heteronuclear systems have been studied \cite{Na-K,Na-Rb,Na-Cs,Rb-Cs,Li-Cs,MOT-single-atom-collision}, aside from collisional processes one has radiative escape. In optical dipole traps there are homonuclear inelastic collisions \cite{collision-ODT-1,collision-ODT-2,collision-ODT-3}, and at sufficiently large densities three-body recombination becomes important \cite{collision-ODT-single}. It is therefore crucial to perform experiments allowing one to single out a particular inelastic process \cite{Ueberholz2000,Fuhrmanek2012,Sompet2013}.

This is done in the present paper. We study a two-atom system of different isotopes of rubidium (single $^{85}$Rb and single $^{87}$Rb) in a micro optical trap. One of them or both are in a higher hyperfine state, and we measure the corresponding rate of inelastic relaxation accompanied by the loss of the atoms. The homonuclear collisions are absent and our measurements give pure loss rates of specific hyperfine heteronuclear collisions. The experiments are done at temperatures close to the border of the ultracold limit (tens of microkelvins) and are supported by finite temperature coupled channels calculations. Our work can be easily extended to other alkali atoms, even to atom-molecule collisions \cite{atom-molecule-collision-1,atom-molecule-collision-2}, thus allowing further understanding of heteronuclear collisions, a precise test of atomic collisional theory, and applications to quantum information processing.

\begin{figure*}[htbp]
\centering
\includegraphics[width=1\linewidth]{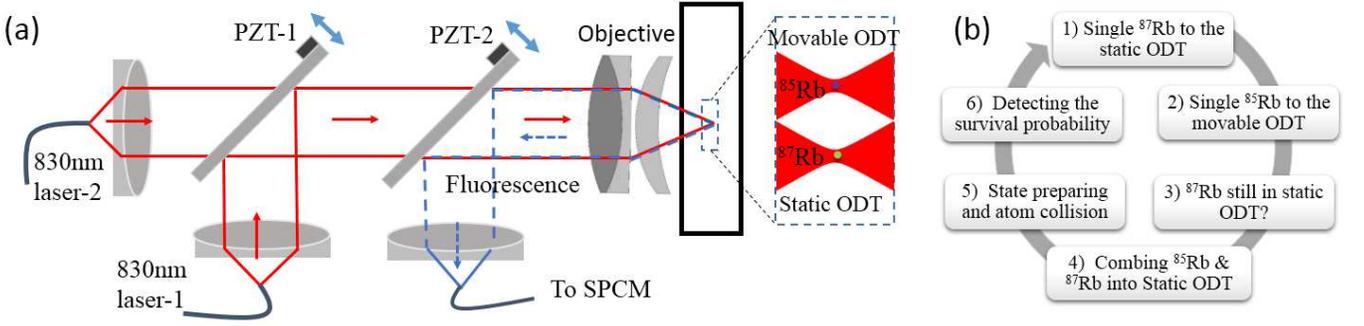}
\caption{(color online) Experimental setup and measurement time sequence. (a) Schematic diagram of the experimental setup. Two 830 nm lasers are collimated and then strongly focused by an objective (Linos,HALO30) into the vacuum chamber to form two ODTs. The movable ODT is from 830 nm laser-1 and can be shifted to overlap with the static ODT (from 830 nm laser-2) by controlling PZT-1 (Piezoelectric Ceramic Transducer). The fluorescence of trapped single atoms is collected by the same objective and guided to SPCM (Single Photon Counting Module) for detection. PZT-2 controls the fluorescence collecting region. A detailed description can be found in Methods. (b)Time sequence in the experiment. Each survival probability in our experiment is the result from 300 repeated measurements.}
\label{fig:fig1}
\end{figure*}

\subsection*{Experimental setup and results}
Our two-atom heteronuclear system is composed of a single $^{85}$Rb and a single $^{87}$Rb in a micro optical dipole trap (ODT), and there are three important points in the experiment. The first one is a sequential trapping of a single $^{87}$Rb in a static ODT and a single $^{85}$Rb in a movable ODT \cite{Beugnon2007}, and we make sure that two atoms of different isotopes are actually trapped (see Fig.\ref{fig:fig1} and Methods). Second, we
shift the movable ODT to overlap with the static one, and adiabatically turn off the movable trap. We get $^{87}$Rb and $^{85}$Rb in one trap with probability of about 95$\%$. The third point is that because of collisional blockade \cite{Collisional-Blockade} we have to kick out one of the atoms before detecting the presence of the other one. By optimizing this procedure we have minimized unwanted atom losses to less than 3$\%$.

Depending on the hyperfine states of $^{87}$Rb and $^{85}$Rb, there are three inelastic decay processes:
\begin{eqnarray*}
(A)\,\,\,\,\,\,\,\,\,\,\,\,\,\,\,\,\,\,^{87}{\rm Rb}(F=2)+^{85}{\rm Rb}(F=3)\Rightarrow\\
\begin{cases}
^{87}{\rm Rb}(F=1)+^{85}{\rm Rb}(F=3)\\
^{87}{\rm Rb}(F=2)+^{85}{\rm Rb}(F=2)\\
^{87}{\rm Rb}(F=1)+^{85}{\rm Rb}(F=2)\end{cases}\\
\end{eqnarray*}
\begin{eqnarray*}
(B)\,\,\,\,\,\,\,\,\,\,\,\,\,\,\,\,\,\,^{87}{\rm Rb}(F=2)+^{85}{\rm Rb}(F=2)\Rightarrow\\
\begin{cases}
^{87}{\rm Rb}(F=1)+^{85}{\rm Rb}(F=3)\\
^{87}{\rm Rb}(F=1)+^{85}{\rm Rb}(F=2)\end{cases}\\
\end{eqnarray*}
\begin{eqnarray*}
(C)\,\,\,\,\,\,\,\,\,\,\,\,\,\,\,\,\,\,&&^{87}{\rm Rb}(F=1)+^{85}{\rm Rb}(F=3)\Rightarrow\\
&&^{87}{\rm Rb}(F=1)+^{85}{\rm Rb}(F=2)
\end{eqnarray*}

We have not set a magnetic field, and for each atomic spin F in the initial state of the collision the states with all possible values of the spin projection $M_{F}$ are likely equally populated. The energy released in the inelastic processes A, B and C is about several GHz and it exceeds the trap depth $U_{0}$ by more than two orders of magnitude. Therefore, both atoms are ejected from the trap as a result of the  inelastic relaxation, which is confirmed in the experiment by selectively kicking out $^{85}$Rb or $^{87}$Rb in the B process. In most of our experiments the trap depth is $U_{0}= 0.6$ mK, which in our configuration provides the radial trap frequency $\omega_{\rho}=38.8\pm0.1$ kHz and the axial trap frequency $\omega_{z}=3.2\pm0.1$ kHz.

\begin{figure*}[htbp]
\centering
\includegraphics[width=0.7\linewidth]{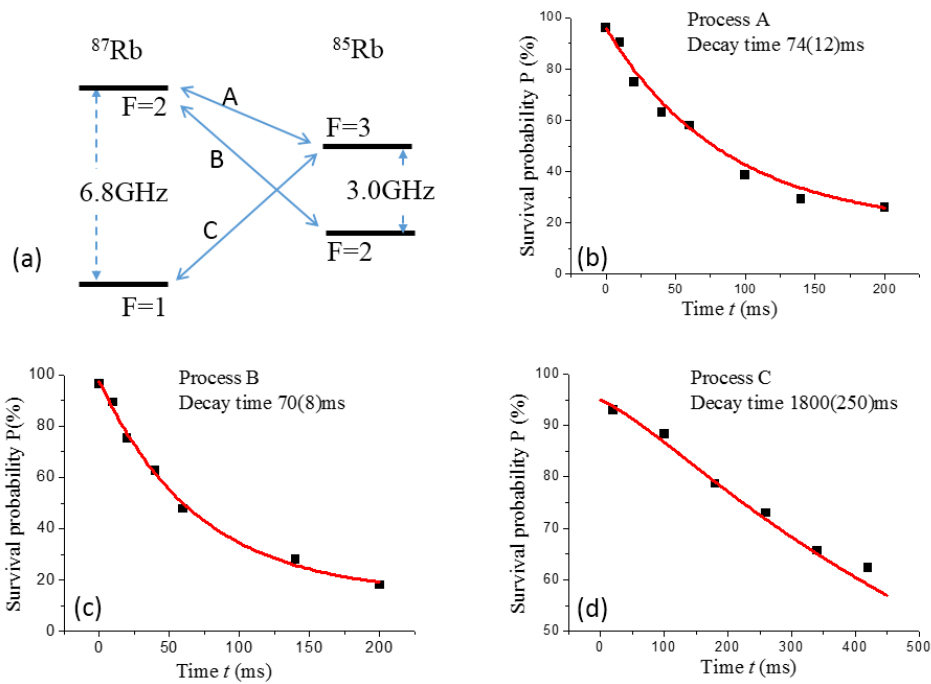}
\caption{(color online) Experimental data for the decay rates. (a) Energy levels of hyperfine states of $^{87}$Rb and $^{85}$Rb. (b), (c), and (d) Survival probability $P$ versus time $t$ for the A, B, and C collisions, respectively. The measurements are done for the survival probability of $^{87}$Rb after kicking out $^{85}$Rb. The black squares are experimental data, with each point being the result from 300 repeated measurements. In b) and c) the solid curves show a fit by the formula $P=w\exp(-t/\tau)+w_{0}$, and in d) a fit with the numerical solution of the rate equations including single atom spin relaxation. The data are collected at the trap depth $U_{0}=0.6$ mK, and the initial temperatures $T_{87}=35\pm3$ $\mu$K, $T_{85}=15\pm1$ $\mu$K. Numbers in bracket are the fitting standard deviations.}
\label{fig:fig2}
\end{figure*}

We measure the survival probability $P(t)$ for the atoms to remain in the trap at time \emph{t} (see Fig.\ref{fig:fig2}). For each \emph{t} we execute 300 repetitions of the loop sequence of Fig.\ref{fig:fig1}(b). In the case of A and B processes the decay is strongly dominated by the interaction-induced spin relaxation. The probability $P(t)$ is then described by an exponential time dependence. Within less than $10\%$ of uncertainty the experimental data can be fitted with an exponential function $P=w\exp(-t/\tau)+w_{0}$. The presence of the residue $w_{0}$ has two reasons. First of all, the two-atom system is obtained with 95$\%$ probability, and there are traps with only one atom which remains trapped on a much longer time scale (about 11s \cite{ourreview}) than the collisional lifetime $\tau$. Second, doubly polarized pairs (for each atom the spin projection is equal to the spin) can decay only due to weak spin-spin or spin-orbit interactions and practically remain stable on the time scale of our experiment.

The process C is much slower than A and B, and our measurements for this process have been made on a time scale of about $500$ ms. In this case the decay is significantly influenced by single atom spin relaxation, and we have to take it into account in the rate equations for extracting $\tau$ from our measurements (see Methods).

The measured $\tau$ has about 15$\%$ of statistical uncertainty which decreases with increasing the executed loop numbers. Aside from single atom spin relaxation, the value of $\tau$  is  influenced by single atom loss events. The heating rate in the dipole trap is about 20 $\mu$K/s \cite{YU2014} and it increases the collisional volume, thus slightly increasing the decay time $\tau$. We estimate the overall uncertainty in our values of $\tau$ as about 20 $\%$.

\begin{figure*}[htbp]
\centering
\includegraphics[width=0.7\linewidth]{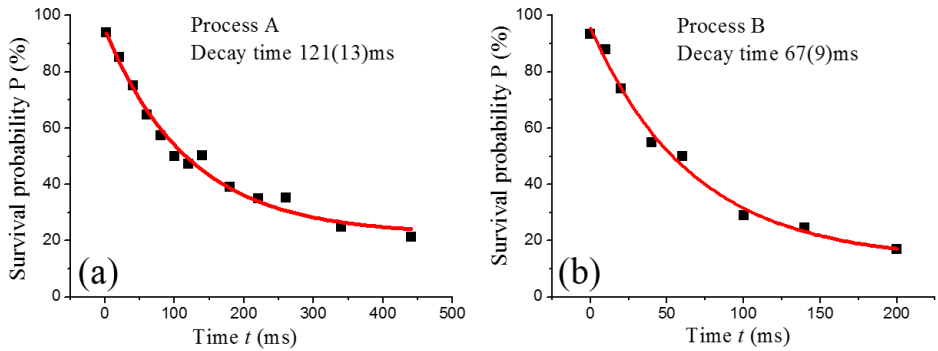}
\caption{(color online) The decay under different conditions. (a) The same as in Fig.\ref{fig:fig2}(b) but for $T_{87}=47\pm3$ $\mu$K and $T_{85}=27\pm2$ $\mu$K. (b) The same as in Fig.\ref{fig:fig2}(c) but for kicking out $^{87}$Rb.}
\label{fig:fig3}
\end{figure*}

For the A collisional process, we also vary the temperature in order to test the dependence of $\tau$ on the effective volume (density) of atoms in the trap. As expected, the time $\tau$ increases with temperature and one can see this from the comparison of the results in Fig.\ref{fig:fig2}(b) and Fig.\ref{fig:fig3}(a). We then tested that the result for $\tau$ does not depend on whether we kick out $^{85}$Rb or $^{87}$Rb for measuring $P(t)$ (see Fig.\ref{fig:fig2}(c) and Fig.\ref{fig:fig3}(b)). It is easy to conclude that not only relaxation times are very close to each other (as one sees in Table 1), but also the functions $P(t)$.

\subsection*{Theory and analysis}
The rate equations for the inelastic decay processes A, B, and C can be written as
\begin{equation}        \label{P}
\frac{dP}{dt}=-\frac{P}{\tau},
\end{equation}
where $P(t)$ is the probability that at time $t$ the atoms are still present in the trap, and $\tau$ is the relaxation (decay) time that we measure. These processes occur at interatomic distances of the order of or smaller than the radius of the interaction potential $R_e=(mC_6/\hbar^2)^{1/4}\approx 80\AA$ ($C_6$ is the Van der Waals constant). At our trap frequencies and temperatures from 15 to 55 $\mu$K we have $T\gg\hbar\omega_{\rho},\hbar\omega_z$, and the motion of atoms in the trap is surely quasiclassical. The extention of the wavefunction of the atoms in the trap is $r_T\sim (T/m\omega^2)^{1/2}\sim 0.2 \mu$m, where $m$ and $\omega$ are characteristic values of the atom mass and trap frequency. Thus, we have the inequality
\begin{equation}      \label{rR}
r_T\gg R_e.
\end{equation}
Therefore, the decay (relaxation) time $\tau$ can be expressed through the relaxation rate constant $\alpha$ in free space:
\begin{equation}           \label{taualpha}
\frac{1}{\tau}=\frac{\alpha}{V_\mathrm{eff}},
\end{equation}
where
\begin{equation}       \label{Veff}
V_\mathrm{eff}=\left(\frac{2\pi T_\mathrm{eff}}{\mu\omega_{\rho}^{4/3}\omega_z^{2/3}}\right)^{3/2}
\end{equation}
is the effective volume, $T_\mathrm{eff}=\mu (T_1/m_1+T_2/m_2)$ is the effective temperature, with $m_1,m_2$ and $T_1,T_2$ being the masses and temperatures of $^{87}$Rb and $^{85}$Rb, and $\mu$ is the reduced mass. The rate constant $\alpha$ in Eq.(\ref{taualpha}) is averaged over the Boltzmann distribution of relative momenta $k$ at effective temperature $T_\mathrm{eff}$
\begin{equation}        \label{alphak}
\!\!\alpha=\!\left(\frac{\hbar^2}{2\pi \mu  T_\mathrm{eff}}\right)^{3/2}\!\!\!\int_0^{\infty}\!\!\!\alpha(k)\exp\!\left(-\frac{\hbar^2k^2}{2\mu T_\mathrm{eff}}\right)4\pi k^2dk.
\end{equation}
Note that due to elastic collisions between the atoms the two-atom system eventually acquires an equilibrium temperature $(T_1+T_2)/2$. However, it is different from the initial effective temperature $T_\mathrm{eff}$ by less than $1\%$, and so will be the effective volume and the average value of $\alpha$.

\begin{table*}
\caption{\label{tab:tab1} Summary of the results (the experimental data are for the trap depth $U_0=0.6$ mK ($\omega_{\rho}=38.8$ kHz, $\omega_z=3.2$ kHz))}
\begin{tabular*}{1\linewidth}{c|c|c|c|c}
    \hline\hline
    \tabincell{c}{\textbf{Collisional} \\ \textbf{process}}   &\tabincell{c}{\textbf{Temperature and} \\ \textbf{trap depth($\mu$K )}} & \textbf{Decay time(ms)}  & \textbf{Experimental $\alpha (cm^3/s)$   }    & \textbf{Calculated $\alpha (cm^3/s)$ }  \\ \hline

     &T$_{87}=35\pm3$, T$_{85}=15\pm1$, kick out $^{85}$Rb       & $\tau_{A}$=74$\pm$12  &  (5.9$\pm$1.1)$\times$ 10$^{-11}$     & 5.6$\times$ 10$^{-11}$     \\ \cline{2-5}

    A  &T$_{87}=47\pm3$, T$_{85}=27\pm2$, kick out $^{85}$Rb       & $\tau_{A}$=121$\pm$13  &  (6.5$\pm$0.8)$\times$ 10$^{-11}$    & 5.9$\times$ 10$^{-11}$    \\\hline

    &T$_{87}=35\pm3$, T$_{85}=15\pm1$, kick out $^{85}$Rb     &  $\tau_{B}$=70$\pm$8 & (6.3$\pm$0.9)$\times$ 10$^{-11}$    \\ \cline{2-4}

    B &T$_{87}=35\pm3$, T$_{85}=15\pm1$, kick out $^{87}$Rb     &  $\tau_{B}$=67$\pm$9 & (6.6$\pm$1.0)$\times$ 10$^{-11}$   & 6.8$\times$ 10$^{-11}$   \\ \hline

    C &T$_{87}=35\pm3$, T$_{85}=15\pm1$, kick out $^{85}$Rb
    & $\tau_{C}$=1800$\pm$250      & (2.4$\pm$0.4)$\times$ 10$^{-12}$                 & 3.2$\times$      10$^{-12}$   \\ \hline

    \hline

\end{tabular*}

\end{table*}

At our temperatures the quantity $k_TR_e\sim 0.5$, where $k_T=(mT/\hbar^2)^{1/2}$ is the thermal momentum, so that we are close to the border of the $s$-wave scattering limit. Therefore, in addition to the $s$-wave scattering, we took into account the scattering with higher orbital angular momenta. The rate constants $\alpha_A,\,\alpha_B$, and $\alpha_C$ for the processes A, B, and C were calculated using the coupled channels method \cite{mole-review,Verhaar} at finite collision energies (see Methods). In the center of mass reference frame the Hamiltonian governing the collisions has the form
\begin{equation}         \label{H}
H=\frac{p^2}{2\mu}+\frac{{\bf \ell}^2}{2\mu r^2}+V_\mathrm{el}(r)+V_\mathrm{hf},
\end{equation}
where $r$ is the interatomic distance, $p$ is its conjugate momentum, and ${\bf \ell}$ is the orbital angular momentum operator. The
interatomic interaction operator is given by $V_\mathrm{el}(r)=V_s(r)P_s+V_t(r)P_t$, with $P_s$ and $P_t$ being projectors onto the electronic singlet and triplet states of the colliding pair of atoms. The term $V_\mathrm{hf}=a_1{\bf S}_1\cdot{\bf I}_1+a_2{\bf S}_2\cdot{\bf I}_2$ is the hyperfine interaction, where ${\bf S}_1,{\bf I}_1,a_1$ and ${\bf S}_2,{\bf I}_2,a_2$ are the electron and nuclear spin operators and hyperfine constants for $^{87}$Rb and $^{85}$Rb, respectively. The total spin operators of the atoms are ${\bf F}_1={\bf S}_1+{\bf I}_1$ and ${\bf F}_2={\bf S}_2+{\bf I}_2$, and the total spin operator of the pair is ${\bf F}={\bf F}_1+{\bf F}_2$. The Hamiltonian $H$ of Eq.(\ref{H}) conserves both the total spin $F$ and its projection $M_F$.
It also conserves the orbital angular momentum $\ell$ and its projection $M_{\ell}$.

The rate of inelastic spin relaxation occuring when at least one of the colliding atoms is in a higher hyperfine state, can be expressed through the real and imaginary parts of the elastic scattering amplitude. The well-known formula for the inelastic rate constant \cite{LL3}, averaged over the initial spin projections, is transformed to (see Methods):
\begin{eqnarray}
&&\alpha(F_1,F_2,k)=\frac{4\pi\hbar}{(2F_1+1)(2F_2+1)\mu}\sum_{\ell=0}^{\infty}(2\ell+1)\sum_{F=|F_1-F_2|}^{F=F_1+F_2}\nonumber  \\
&&\times (2F+1)[{\rm Im}f_{\ell}(F_1,F_2,F,k)-
k|f_{\ell}(F_1,F_2,F,k)|^2].   \label{alphaF}
\end{eqnarray}
Accordingly, $\alpha_A\equiv\alpha(2,3),\,\alpha_B\equiv\alpha(2,2)$, and $\alpha_C\equiv\alpha(1,3)$. The quantity
$f_{\ell}(F_1,F_2,F,k)$ in Eq.(\ref{alphaF}) is the amplitude of elastic ${\ell}$-wave scattering of these atoms at the total spin $F$.

We apply the accumulated--phase method (see Methods) and calculate the accumulated--phase parameters
from the known properties of homonuclear ${}^{87}\mathrm{Rb}{}^{87}\mathrm{Rb}$ and ${}^{85}\mathrm{Rb}{}^{85}\mathrm{Rb}$ collisions using
mass scaling \cite{Verhaar}.
The main inaccuracy of our calculations stems from the choice of the accumulated phase, and we have checked that our results
are stable within 5\% when the value of this phase is varied by a few percent.
The $p$-wave ($\ell=1$) contribution at our temperatures is comparable with the $s$-wave ($\ell=0$) one,
but the contributions of the $d$-wave and higher partial waves are below $1\%$.
Therefore, in the following we confine ourselves only to the $s$-wave and $p$-wave scattering.

In Table 1 the results of the calculations for the processes A, B and C are compared with the experimental data. For the fast A and B  processes one sees an agreement between experiment and theory within the error bars of the experimental data. For the slow C process the calculated $\alpha$ is near the upper bound of the experimental value accounting only for statistical uncertainties. The reason for this small discrepancy is that the heating effect, although fairly small, still increases the effective volume so that the measured $\tau$ actually corresponds to slightly higher temperatures than the initial ones. This means that the experimental value of $\alpha$ at the initial temperatures should actually be slightly (by about $15\%$) higher than the one in Table 1.

\subsection*{Discussion}
The relaxation rates obtained in our work are rather high. In interesting experiments with spinor heteronuclear mixtures \cite{Papp2008}, this places an upper limit of at about $n\sim 10^{12}$ cm$^{-3}$ on the density if a higher hyperfine atomic state is involved. Our relaxation rate constants are several orders of magnitude higher than the ones for doubly polarized atomic states, like for example $^{87}$Rb$(F_1=2,M_1=2)$, where the relaxation is caused only by weak spin-spin and spin-orbit interactions \cite{chin-review}. We expect a similar reduction for inelastic collisions of doubly polarized $^{87}$Rb and $^{85}$Rb atoms, such as $^{87}$Rb$(F_1=1,M_1=1)$+$^{85}$Rb$(F_2=3,M_2=3)$.

Our results may also have implications for engineering of quantum states and creation of two-qubit quantum gates in a system of single-atom traps, in particular with respect to a proper selection of hyperfine atomic states.
The relaxation time $\tau$ in our experiment is still much larger than the characteristic time on which the single-atom coherence is preserved - the decoherence time $\tau_{dc}$. The time $\tau_{dc}$ measured at the same temperature $T$ is about 1 ms and it increases with decreasing $T$ \cite{Kuhr2005,Yushi2013}. On the other hand, the time $\tau$ strongly decreases with $T$ due to a related decrease in the effective volume $V_\mathrm{eff}$. The expected reduction in $\tau$ for the atoms in the ground vibrational state is about a factor of 30 compared to the data in Table 1, and it becomes of the same order of magnitude as  $\tau_{dc}$, at least for the atoms in the states $F_1=2, F_2=2$ and $F_1=2,F_2=3$ (processes A and B). Even for the slowest process C the expected $\tau$ is $\sim 50$ms, which is close to the maximum $\tau_{dc}$ obtained up to now in experiments with single-atom traps \cite{li2012}.

The time $\tau$ can be certainly increased by decreasing the trap frequencies. However, there is a lower bound for the frequencies. This is because for quantum gates the time $\tau_{dc}$ and, hence, $\tau$ should be at least 4 orders of magnitude larger than the operational time $\tau_{op}$ \cite{Divicenzo 2000}. The latter is on the millisecond level in the considered situation and can not be much smaller than the inverse interaction energy of the two atoms in the trap, so that $\tau_{op}\sim \hbar V_\mathrm{eff}/g$, where $g=4\pi\hbar^2a/m$ is the coupling constant for the elastic interaction, and $a$ is the scattering length. We thus have the condition:
\begin{equation}       \label{condop}
\tau\gg\tau_{dc}>10^4\tau_{op}\sim \frac{10^4\hbar V_\mathrm{eff}}{g}.
\end{equation}
As the effective volume is $V_\mathrm{eff}\propto 1/\omega^3$, for realistic $\tau_{dc}$ below 1s the lower bound for the trap frequency is about a few kiloHertz.

There is also a fundamental limitation independent of the trap frequencies. Since $\tau=V_\mathrm{eff}/\alpha$, equation (\ref{condop}) immediately leads to the inequality
\begin{equation}       \label{limitgalpha}
\frac{g}{\hbar\alpha}\gg 10^4.
\end{equation}
For common values of $g/\hbar$ from $10^{-9}$ to $10^{-11}$ cm$^3$/s equation (\ref{limitgalpha}) can be satisfied for doubly polarized atomic states, such as $^{87}$Rb$(F=2,M_F=2)$ where $g/\hbar\sim 10^{-10}$ cm$^3$/s, and $\alpha<10^{-14}$ cm$^3$/s. Another option would be to increase $g$ by using a Feshbach resonance, although this can also increase inelastic losses.

The advantage of our work is that we study collisions in a two-atom system, which allows us to single out a particular collisional process. In upcoming experiments we intend to prepare atomic states with given spin projections $M_F$ and cool the atoms down to the ground vibrational state \cite{Kaufman2012,Thompson2013} in order to execute a possibility of creating a quantum gate. Our work also paves a way to the creation of single heteronuclear molecules and to the studies of atom-molecule and molecule-molecule binary systems.

\subsection*{Methods}
{\it Details of experimental methods}\\
In our experiment single atoms are trapped in micro optical dipole traps (ODT) which are formed by strongly focusing 830 nm lasers to a beam waist of about 2.1 $\mu$m. The dipole lasers follow the paths shown in Fig.\ref{fig:fig1}(a) by the red solid lines. The movable ODT is initially 5 $\mu$m away,
and it can be shifted to overlap with the static ODT by changing the voltage of PZT-1. We detect trapped atoms by collecting the fluorescence in the trap region as shown by blue dashed lines. The fluorescence is coupled into a polarization-maintaining (PM) single mode fiber (with the core diameter of 5 $\mu$m) for spatial filtering and is then guided to a single photon counting mode (SPCM, AQRH-14-FC). Owing to a moderate core diameter of the fiber and to a fairly large distance between the two traps, we can selectively collect the fluorescence from one of them. The disturbance from the other trap (crosstalk effect) is eliminated by properly adjusting the voltage of PZT-2.

The whole experiment is executed following the time sequence shown in Fig.\ref{fig:fig1}(b). We first trap $^{87}$Rb in the static ODT and then $^{85}$Rb in the movable trap. We do an extra check of the $^{87}$Rb presence in the trap by turning on the $^{87}$Rb MOT again for 20ms at step 3. Only when $^{87}$Rb is still trapped, we record the final result of the collision. In order to check that only a $^{87}$Rb is in the static trap and meanwhile only a $^{85}$Rb in the movable trap, we first detect the absence of the fluorescence of $^{87}$Rb from the movable trap and the absence of fluorescence of $^{85}$Rb from the static trap. This is followed by the fluorescence detection of $^{87}$Rb and $^{85}$Rb in the static and movable traps, respectively.

In step 4 we first prepare $^{85}$Rb in $F_2=2$ and $^{87}$Rb in $F_1=1$ states in order to eliminate the unwanted collisional losses when switching off the MOT repumping lasers 1ms before the MOT cooling lasers. For minimizing the heating effect we have optimized the process of transferring $^{85}$Rb to the static trap. The transfer efficiency can go up to 98\% and is limited by the detection efficiency and heating losses.
The probability that $^{87}$Rb survives when the movable trap approaches the static one and then is adiabatically switched off, is also about 98\%. The temperature is measured in this type of process by using the release and recapture technique \cite{ourreview} when one of the traps (either static or movable) is empty.

In this way we create a heteronuclear two-atom system. After a certain time, we kick out one of the atoms from the trap by using resonant lasers. Optimizing the laser intensities and shortening the pulse duration to 0.1ms, we have minimized the unwanted losses to less than 3\%. Eventually, we have succeeded in trapping two heteronuclear single atoms in the static ODT with about 95$\%$ probability.\\

{\it Analysis of the experimental data}\\
For the slow process C we extracted $\tau$ from the fit of our experimental data with the numerical solution of the rate equations taking into account single atom spin relaxation. These are linear equations for the quantities $P_{F_1F_2}(t)$ representing the probabilities that both $^{87}$Rb with spin $F_1$ and  $^{85}$Rb with spin $F_2$ are present in the trap at time $t$:
\begin{eqnarray}
&&\frac{dP_{13}}{dt}=-\frac{P_{13}}{\tau_C}+\frac{(P_{23}+P_{12}-2P_{13})}{\tau_r}, \nonumber \\
&&\frac{dP_{12}}{dt}=\frac{(P_{13}+P_{22}-2P_{12})}{\tau_r}, \nonumber \\
&&\frac{dP_{23}}{dt}=-\frac{P_{23}}{\tau_A}+\frac{(P_{13}+P_{22}-2P_{23})}{\tau_r}, \nonumber \\
&&\frac{dP_{22}}{dt}=-\frac{P_{22}}{\tau_B}+\frac{(P_{23}+P_{12}-2P_{22})}{\tau_r}. \nonumber
\end{eqnarray}
For the time of single atom spin relaxation we did an independent measurement, and the measured value is $\tau_r=1100\pm150$ ms. The interaction-induced relaxation times $\tau_A$ and $\tau_B$ were taken from the data for the A and B processes. In the main text, Fig.\ref{fig:fig2}(d) shows that the best numerical fit $P_{13}(t)$ deviates from the experimental data by less than $10\%$.\\

{\it Calculation of the inelastic rate constants}  \\
At low temperature the inelastic spin relaxation occurs when at least one of the colliding atoms is in a higher hyperfine state. The rate constant of this process is given by \cite{LL3}
\begin{equation}        \label{alpha1}
\!\!\alpha(F_1,F_2,k)\!=\!\frac{\pi\hbar}{\mu k}\!\sum_{M_1,M_2,f}\sum_{\ell=0}^{\infty}\frac{(2\ell+1)|S_{if}^{(\ell)}|^2}{(2F_1+1)(2F_2+1)}.
\end{equation}
The quantity $S_{if}^{(\ell)}$ is the $S$-matrix element for the $\ell$-wave scattering from the initial state $i$ characterized by the atom spins $F_1,F_2$ and their projections $M_1,M_2$ to a final state $f$ which has a lower internal (hyperfine) energy, so that there is an energy release in the inelastic scattering process. In Eq.(\ref{alpha1}) we also averaged over the initial spin projections $M_1$ and $M_2$. Due to the unitarity condition for the $S$-matrix elements we have:
\begin{equation}          \label{unitarity}
\sum_f|S_{if}^{(\ell)}|^2=1-|S_{ii}^{(\ell)}|^2-\sum_{i'\neq i}|S_{ii'}^{(\ell)}|^2,
\end{equation}
where $S_{ii}$ is the $S$-matrix element for elastic scattering in which the spin projections $M_1,M_2$ remain the same,
and $S_{ii'}$ is the $S$-matrix element for elastic scattering which changes $M_1,M_2$ to $M_1^{\prime},M_2^{\prime}$.
Expressions for the $S$--matrix elements through the corresponding scattering amplitudes read \cite{LL3}:
\begin{equation} \label{eq:Sf}
  S_{ii'}^{(\ell)}=\delta_{ii'}+2ikf_{ii'}^{(\ell)}(k).
\end{equation}
The amplitudes $f_{ii'}$ are conveniently expressed through the amplitudes
$f_{\ell}(F_1,F_2,F,k)$ of elastic scattering of atoms with spins $F_1,F_2$ at the total spin $F$ of the pair:
\begin{eqnarray}
&&f_{ii'}^{(\ell)}(k)=\sum_{F=|F_1-F_2|}^{F=F_1+F_2}\langle F_1M_1F_2M_2|F_1F_2FM\rangle  \nonumber \\
&&\langle F_1M_1^{\prime}F_2M_2^{\prime}|F_1F_2FM\rangle f_{\ell}(F_1,F_2,F,k).\label{ff}
\end{eqnarray}
In the absence of a magnetic field the amplitudes $f_{\ell}(F_1,F_2,F,k)$ do not depend on the total spin projection $M$.
The Clebsch--Gordan coefficients which appear in Eq.(\ref{ff}) satisfy the summation rule:
\begin{eqnarray}
&&\sum_{M_1,M_2} \langle F_1M_1F_2M_2|F_1F_2FM\rangle \langle F_1M_1F_2M_2|F_1F_2\tilde F\tilde M\rangle\  \nonumber \\
&&=\delta_{F\tilde F}\delta_{M\tilde M}.\label{summation}
\end{eqnarray}
Substituting Eqs.(\ref{eq:Sf}) and (\ref{ff}) into Eq.(\ref{alpha1}), and making use of Eq.(\ref{summation}),
we arrive at equation (\ref{alphaF}) of the main text:
\begin{multline}
\alpha(F_1,F_2,k)=
\frac{1}{(2F_1+1)(2F_2+1)}\:
\frac{4\pi\hbar}{\mu} \:
\sum_{\ell=0}^{\infty}(2\ell+1)\\
\sum_{F=|F_1-F_2|}^{F_1+F_2}(2F+1)
[\mathrm{Im}f_{\ell}(F_1,F_2,F,k)-
k|f_{\ell}(F_1,F_2,F,k)|^2]. \nonumber
\end{multline}

The amplitudes $f_{\ell}(F_1,F_2,F,k)$ were calculated numerically using the coupled channels method \cite{mole-review}.
Our implementation of this method is described in Ref. \cite{PapoularPhD2011}.
The asymptotic behavior of the scattering states is enforced at a distance of $1000a_0$,
 with $a_0$ being the Bohr radius. The accumulated-phase boundary condition is applied at $r_0=16a_0$.
It summarizes the short--range physics in the region of distances $r<r_0$,
where the triplet and singlet interaction potentials are poorly known, into $6$ phase parameters.
We calculate these parameters for heteronuclear ${}^{87}\mathrm{Rb}{}^{85}\mathrm{Rb}$ collisions starting
from the known data for homonuclear ${}^{87}\mathrm{Rb}{}^{87}\mathrm{Rb}$ and ${}^{85}\mathrm{Rb}{}^{85}\mathrm{Rb}$
collisions and using the  mass scaling technique \cite{Verhaar}, which exploits the fact
that the Born--Oppenheimer electronic potentials $V_s$ and $V_t$ do not depend on the type of isotopes.
The hyperfine coefficients $a_1$ and $a_2$ were taken from Ref. \cite{arimondo:RMP1977}.
Solving the coupled differential equations for the wavefunctions within the subspaces characterized by the conserved quantum numbers $F,M$
we calculated all scattering amplitudes $f_{\ell}(F_1,F_2,F,k)$ as functions of the incident relative momentum $k$.
The calculated rate constants $\alpha(F_1,F_2,k)$ are then averaged over the thermal distribution of
$k$ according to Eq.(\ref{alphak}). Note that our zero temperature result $\alpha_C=0.8\times 10^{-12}$ cm$^3$/s is fairly close to the lower bound of the interval $(1.2 - 4.5)\times 10^{-12}$ cm$^3$/s found for the C process with $^{87}$Rb$(F_1=1,M_1=-1)$ and $^{85}$Rb$(F_2=3,M_2=3)$ in Ref. \cite{Bohn} which used old data for the interaction potentials.

\subsection*{Acknowledgments}
We acknowledge fruitful discussions with Jiaming Li, Antoine Browaeys and Philippe Grangier. This work has been supported by the National Basic Research Program of China under Grant No. 2012CB922101 and the National Natural Science Foundation of China under Grant Nos. 11104320 and 11104321. GVS acknowledges support from IFRAF and from the Dutch Foundation FOM. DJP and GVS emphasize that the research leading to their results in this paper has received funding from the European Research Council under European Community's Seventh Framework Programme (FR7/2007-2013 Grant Agreement no.341197).

\noindent
{\bf Author Contributions}
X.D. H., P. X. and M.S. Z. designed the experiments. M.S.Z supervised the project.
J.H. Y, Y. Z. and K.P. W. did the experiment. P.X. and J.W. analyzed the experimental data.
D.J.P., G.V.S., and M.L. performed the theoretical calculation. P. X., G.V. S. and M.S. Z. wrote the paper.

\noindent
{\bf  Competing financial interests}  The authors declare no competing financial interests.

\noindent
{\bf  Corresponding author} Correspondence should be addressed to Mingsheng Zhan (mszhan@wipm.ac.cn).

\end{document}